\documentclass[twocolumn]{aastex61}
\submitjournal{ApJ}

\usepackage{hyperref}
\usepackage{amsmath}
\usepackage{amssymb}
\usepackage{graphicx}
\usepackage[utf8]{inputenc}

\newcommand{\kms}{km\,s$^{-1}$}
\newcommand{\Teff}{$T_{\rm eff}$}
\newcommand{\logg}{$\log g$}

\newcommand{\rproc}{$r$-process }

\newcommand{\ncap}{neutron-capture }
\newcommand{\AB}[2]{$\mbox{[#1/#2]}$}
\newcommand{\feh}{\AB{Fe}{H}}
\newcommand{\xh}{\AB{X}{H}}
\newcommand{\xfe}{\AB{X}{Fe}}

\newcommand{\rI}{$r$-I }
\newcommand{\rII}{$r$-II }
\newcommand{\vmic}{$v_\textrm{micr}$}
\usepackage[utf8]{inputenc}
\shorttitle{Chemical Abundances for a Trio of $R$-Process-Enhanced Stars}
\shortauthors{Cain et al.}
\begin{document}

\title{The $R$-Process Alliance: Chemical Abundances for a Trio of
$r$-Process-Enhanced Stars -- One Strong, One Moderate, and One Mild 
\footnote{This paper includes data gathered with the 6.5\,m Magellan Telescopes located at Las Campanas Observatory, Chile. 
Based on observations at Kitt Peak National Observatory, National Optical Astronomy Observatory (NOAO Prop. IDs: 14A-0323), which is operated by the Association of Universities for Research in Astronomy (AURA) under cooperative
agreement with the National Science Foundation. The authors are honored to be permitted to conduct astronomical research on Iolkam Du'ag (Kitt Peak), a mountain with particular significance to the Tohono O'odham. 
Based on observations obtained at the Gemini Observatory (Prop. IDs:
GS-2015A-Q-205), which is operated by the Association of Universities for
Research in Astronomy, Inc., under a cooperative agreement with the NSF on
behalf of the Gemini partnership: the National Science Foundation (United
States), the National Research Council (Canada), CONICYT (Chile), Ministerio de
Ciencia, Tecnolog\'{i}a e Innovaci\'{o}n Productiva (Argentina), and
Minist\'{e}rio da Ci\^{e}ncia, Tecnologia e Inova\c{c}\~{a}o (Brazil).
Based on observations collected at the European Organisation for Astronomical
Research in the Southern Hemisphere under ESO programme(s) 092.D-0308(A).
}}
\correspondingauthor{Anna Frebel}
\email{afrebel@mit.edu}
\author{Madelyn Cain}
\affiliation{Department of Physics \& Kavli Institute for Astrophysics and Space Research, Massachusetts Institute of Technology, Cambridge, MA 02139, USA}
\author{Anna Frebel}
\affiliation{Department of Physics \& Kavli Institute for Astrophysics and Space Research, Massachusetts Institute of Technology, Cambridge, MA 02139, USA}
\affiliation{Joint Institute for Nuclear Astrophysics - Center for Evolution of the Elements, USA}
\author{Maude Gull}
\affiliation{Department of Physics \& Kavli Institute for Astrophysics and Space Research, Massachusetts Institute of Technology, Cambridge, MA 02139, USA}
\author{Alexander P. Ji}
\affiliation{The Observatories of the Carnegie Institution of Washington, Pasadena, CA 91101, USA}
\affiliation{Hubble Fellow}
\author{Vinicius M. Placco}
\altaffiliation{Visiting astronomer, Kitt Peak National Observatory.}
\affiliation{Department of Physics, University of Notre Dame, Notre Dame, IN 46556, USA}
\affiliation{Joint Institute for Nuclear Astrophysics - Center for Evolution of the Elements, USA}
\author{Timothy C. Beers}
\affiliation{Department of Physics, University of Notre Dame, Notre Dame, IN 46556, USA}
\affiliation{Joint Institute for Nuclear Astrophysics - Center for Evolution of the Elements,  USA}
\author{Jorge Mel\'endez}
\affiliation{Instituto de Astronomia,  Geof\'{i}sica e Ci\^{e}ncias Atmosf\'{e}ricas, 
Universidade de S\~{a}o Paulo, SP 05508-900, Brazil}
\author{Rana Ezzeddine}
\affiliation{Joint Institute for Nuclear Astrophysics - Center for Evolution of the Elements, USA}
\affiliation{Department of Physics \& Kavli Institute for Astrophysics and Space Research, Massachusetts Institute of Technology, Cambridge, MA 02139, USA}
\author{Andrew R. Casey}
\affiliation{Monash Centre for Astrophysics, School of Physics \& Astronomy, Monash University, Clayton, Melbourne 3800, Australia}
\author{Terese T. Hansen}
\affiliation{The Observatories of the Carnegie Institution of Washington, 813 Santa Barbara St., Pasadena, CA 91101, USA}
\author{Ian U.\ Roederer}
\affiliation{Department of Astronomy, University of Michigan, Ann Arbor, MI 48109, USA}
\affiliation{Joint Institute for Nuclear Astrophysics - Center for Evolution of the Elements, USA}
\author{Charli Sakari}
\affiliation{Department of Astronomy, University of Washington, Seattle, WA 98195-1580, USA}
\affiliation{Joint Institute for Nuclear Astrophysics - Center for Evolution of the Elements, USA}

\begin{abstract}

We present detailed chemical abundances of three new bright (V $\sim 11$), extremely metal-poor ([Fe/H] $\sim -3.0$), $r$-process-enhanced halo red giants based on high-resolution, high-$S/N$ Magellan/MIKE spectra. We measured abundances for 20-25 neutron-capture elements in each of our stars. J1432$-$4125 is among the most \rproc rich \rII stars, with [Eu/Fe$]= +1.44\pm0.11$.  J2005$-$3057 is an \rI star with [Eu/Fe$] = +0.94\pm0.07$. J0858$-$0809 has [Eu/Fe$] = +0.23\pm0.05$ and exhibits a carbon abundance corrected for evolutionary status of $\mbox{[C/Fe]}_{\rm{corr}} = +0.76$, thus adding to the small number of known carbon-enhanced $r$-process stars. All three stars show remarkable agreement with the scaled solar \rproc pattern for elements above Ba, consistent with enrichment of the birth gas cloud by a neutron star merger. The abundances for Sr, Y, and Zr, however, deviate from the scaled solar pattern. This indicates that more than one distinct \rproc site might be responsible for the observed neutron-capture element abundance pattern. Thorium was detected in J1432$-$4125 and J2005$-$3057. Age estimates for J1432$-$4125 and J2005$-$3057 were adopted from one of two sets of initial production ratios each by assuming the stars are old. This yielded individual ages of $12\pm6$\,Gyr and $10\pm6$\,Gyr, respectively. 
\end{abstract}


\section{Introduction}\label{intro}
The chemical composition of the oldest stars provides key information about the evolution of  elements in the early universe. Ideal candidates for study are long-lived, very metal-poor ([Fe/H] $< -2.0$) and extremely metal-poor ([Fe/H] $< -3.0$) stars. These stars are believed to have formed from gas enriched by only a few progenitor supernovae or nucleosynthetic events \citep{Frebel15}. Metal-poor stars that are highly enhanced with heavy elements ($Z>30$) are of particular interest, as they are valuable in tracing the yields of heavy element
production events in the early universe. Elements heavier than zinc are built up by neutron-capture in two primary processes 
\citep{frebel18}, the slow neutron-capture process ($s$-process) and the rapid neutron-capture process ($r$-process). Other processes 
such as the $i$-process (\citealt{dardelet14, hampel16, clarkson18}) may also contribute to the formation of \ncap elements. The $r$-process, first described in \citet{Burbidge57} and \citet{cameron57}, is responsible for producing the heaviest elements in our universe. During the $r$-process, seed nuclei (e.g., C or Fe) are rapidly bombarded with neutrons  
to create heavy elements up to and including uranium \citep{Sneden08}. 

A small fraction (3--5\%) of known metal-poor stars with $\mbox{[Fe/H]} <-2.5$ appear to have formed from gas enriched by a previous $r$-process event \citep{Barklem05}. These stars exhibit the same distinct $r$-process chemical pattern as found in the Sun, characterized by three peaks of elements (Se, Br and Kr; Te, I, and Xe; and Os, Ir and Pt) 
to which unstable isotopes produced during the \rproc most frequently decay. Examples of known metal-poor \rproc stars include CS~22892-052 \citep{Snedenetal:1996}, CS~31082-001 \citep{Hill02}, HE~1523$-$0901 \citep{Frebel07}, and others \citep{Barklem05,placco17, sakari18, holmbeck18}. The chemical signatures of these stars are identical to the scaled solar \rproc abundance pattern for elements with $Z\geq 56$, indicating that the $r$-process pattern is in fact universal for Ba and above (with the exception of the actinides). For the lighter elements ($38\leq Z<56$), variations exist between the patterns found in metal-poor stars and the Sun (e.g., \citealt{Barklem05, Roederer14d,Ji16b,ji18}). 

The discrepancies in the \rproc abundance pattern for $38 \leq Z < 56$ indicate that the overall \rproc pattern may be a composite of two processes: the limited $r$-process (sometimes called the weak $r$-process), where the neutron flux is too low to produce elements beyond the second \rproc peak in significant quantities \citep{Truran02}, and the main $r$-process, which primarily populates the second peak and beyond \citep{frebel18}. The limited \rproc ([Eu/Fe$]<+0.30$, [Sr/Ba$]>+0.50$, [Sr/Eu$]>+0.00$) is thought to produce elements with $38 \leq Z < 56$ in higher quantities compared to the heavier elements. So far, the \ncap element 
poor stars HD~88609 and HD~122563 are the best candidates stars formed from gas solely enriched by the limited $r$-process (\citealt{Honda07,honda2007}).  The main \rproc results in the characteristic $r$-process pattern for elements Ba and above, excluding the actinides. Stars that exhibit this pattern are categorized as either moderately \ncap 
enhanced \rI stars ($+0.3 <$ [Eu/Fe] $\leq +1.0$) or strongly enhanced $r$-II stars ([Eu/Fe$]> +1.0$) \citep{Beers05}. Currently, only $\sim 30$-$40$ \rII stars and  $\sim 125$-$150$ \rI stars have been recognized. 

There are many open questions regarding the astrophysical site(s) of the $r$-process. The main \rproc is now firmly believed to occur during the mergers of binary neutron stars or a neutron star and a black hole. Theoretically, this has long been suggested  (\citealt{Lattimer74, freiburghaus99, rosswog14}). Recently, the discovery of the \rproc ultra-faint dwarf galaxy Reticulum\,II \citep{Ji16a,Roederer16a} has provided observational support for this interpretation. Since it was possible to estimate the gas mass into which the $r$-process yield was diluted in Reticulum\,II, available yield predictions could be compared with the observed stellar abundances for the first time. Good agreement between the diluted yield and measured abundances provided convincing evidence for enrichment by a rare and prolific $r$-process event in the early universe, consistent with a neutron star merger. 

The detection of local kilonova transients \citep{Tanvir13} further confirms the production of \ncap elements during neutron star mergers. Recently, the LIGO – Virgo gravitational-wave  observatory network detected the merger of a neutron star pair, GW170817 \citep{LIGOGW170817a,LIGOGW170817b}. The associated electromagnetic counterpart, SSS17a, revealed a kilonova transient in the dynamical ejecta and post-merger winds following the production of \ncap elements (\citealt{coulter17, drout17, kilpatrick17, shappee17}). Additional evidence for neutron star mergers as the site of the main $r$-process comes from Pu measurements from deep-sea ice cores, which suggest that the local universe is primarily enriched by a rare and massive \rproc event, rather than multiple smaller enrichments by core-collapse supernovae \citep{Hotokezaka15, Wallner15}.
  
On the contrary, the limited \rproc is likely active in other astrophysical sites. Evidence from Galactic halo stars indicates that it may occur during core-collapse supernovae, possibly through a high-entropy neutrino wind (e.g., \citealt{Meyer92,Woosley92, Kratz07,Arcones11,Wanajo13}) or a rotating proto-neutron star (e.g., \citealt{Cameron03, Winteler12, Nishimura15}). 

To further study the \rproc and its astrophysical site(s), it is valuable to study \rproc stars  in the Galactic halo. These stars likely formed in small dwarf galaxies in the early universe from gas enriched by one or few \rproc events. Analyzing metal-poor \rproc halo stars provides one clear advantage over dwarf galaxy stars. Namely, they are bright 
and can be easily observed to obtain a very high $S/N$ spectrum necessary for a detailed abundance analysis.  In this paper, we present a newly discovered \rII star, 2MASS~J14325334$-$4125494 (hereafter J1432$-$4125), and one \rI star, 2MASS~J20050670$-$3057445 (hereafter J2005$-$3057). We also analyze 2MASS~J08580584$-$ 0809174 (hereafter J0858$-$0809), a mildly \rproc enhanced star CEMP star. These stars were found as part of the ongoing work of the $R$-Process Alliance \citep[RPA;][]{hansen18}, a recently formed collaboration that aims to combine observations, theory and modeling, and experiments from both astrophysics and nuclear physics to advance our knowledge of the $r$-process. We now present the first detailed abundance analysis of J0858$-$0809, J1432$-$4125, or J2005$-$3057 with high resolution spectra. The chemical abundance data for these stars will provide  deeper insight to the \rproc chemical abundance pattern and the production site(s) of the $r$-process. 

\section{Observation and Line Measurements}\label{target}

\begin{deluxetable*}{lrrccccrccccr}[hbt!]
\tablecaption{\label{table:observations} Observation Details }
\tablehead{               
\colhead{Star} & \colhead{$\alpha$}&\colhead{$\delta$}&\colhead{UT dates}&\colhead{UT times}&\colhead{slit}&\colhead{$t_{\rm {exp}}$}
&\colhead{V}  & \colhead{B$-$V} &\colhead{$S/N$}  &\colhead{$S/N$} &\colhead{$S/N$} &\colhead{$\textrm{v}_{\textrm{helio}}$}   \\
\colhead{}&\colhead{[J2000]}&\colhead{[J2000]}&\colhead{}&\colhead{}&\colhead{}&\colhead{[min]}&\colhead{[mag]}&\colhead{[mag]}&\colhead{[4000\,{\AA}]}&\colhead{[4500\,{\AA}]}&\colhead{[6000\,{\AA}]
}&\colhead{[\kms]}} 
\startdata 
J0858$-$0809& 08 58 05.8 &$-$08 09 17  & 2016-04-15 & 01:33:45 &$0\farcs7$ & 10.0&  10.49 & 0.61 & 170 &265& 440& +168.1\\
&  &  & 2016-04-17  &   23:31:11 &$0\farcs7$ & 30.0&  & & & & &+170.8\\
J1432$-$4125& 14 32 53.3 &$-$41 25 49  & 2016-04-16 & 03:50:54 &$0\farcs7$ & 10.0&  11.10& 0.60 & 105 & 170& 220&$-228.7$ \\
J2005$-$3057& 20 05 06.6 &$-$30 57 44  & 2016-04-16  & 06:54:49 &$0\farcs7$ & 30.0& 11.80 & 0.60 & 75 & 185& 320&$-265.9$ \\
\enddata 
\tablecomments{$S/N$ is per pixel. J0858$-$0809 was observed over two nights, so we list the $S/N$ value that result from the combined spectrum from both nights. } 
\label{tabel:observations}
\end{deluxetable*}

\subsection{Target Selection and Observations}
J0858$-$0809 and J2005$-$3057 were first identified as metal-poor star candidates in the RAVE DR5 database \citep{kunder17} using selection techniques described in \citet{Placco18}. They were then followed up with medium-resolution spectroscopy using the KPNO/Mayall (RC Spectrograph - semester 2014A) and Gemini South (GMOS - semester 2015A) telescopes, respectively.
J1432$-$4125 was first identified as a potential low-metallicity star from its photometry, based on the methods described in \citet{melendez16}. It was then followed-up with medium-resolution spectroscopy in semester 2014A using the
EFOSC-2 spectrograph at the ESO New Technology Telescope. The observing setup was similar for all three telescope/spectrograph combinations. We used gratings ($\sim$600~l~mm$^{\rm{-1}}$) and slits ($\sim 1\farcs$0) in the blue setup, covering the wavelength range $\sim$3550-5500\,{\AA}. This combination yielded a resolving power of $R\sim 2,000$, and exposure times were set to yield signal-to-noise ratios of S/N $\sim 50$ per pixel at 3900\,{\AA}. 
The calibration frames included arc-lamp exposures (taken following the science observations), bias frames, and quartz-lamp flatfields. Calibration and extraction were performed using standard IRAF\footnote{\href{http://iraf.noao.edu} {http://iraf.noao.edu}.} packages. Further details on the medium-resolution spectroscopy observations are provided in \citet{Placco18}. Stellar atmospheric parameters and carbon abundances were determined from the
medium-resolution spectra, using the n-SSPP pipeline \citep{beers14,beers17}. The stellar parameters obtained are listed in Table \ref{table:parameters}. The estimated [C/Fe] abundances were $+0.21$, $+0.49$, and $-0.30$ for J0858$-$0809, J1432$-$4125, and J2005$-$3057, respectively.

\begin{figure*}[!ht]
\begin{center}
  \includegraphics[clip=false,width=.9\textwidth]{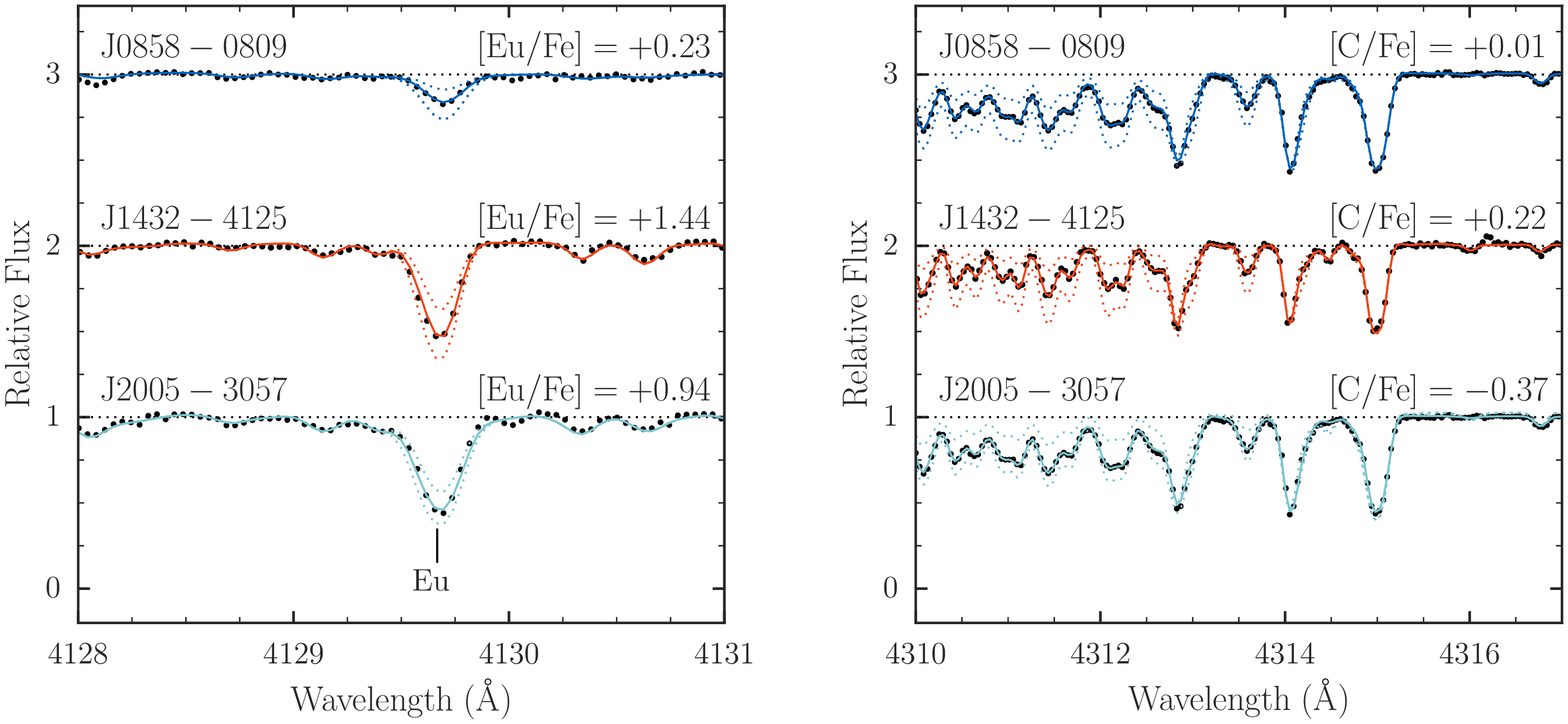} 
  \caption{
     \label{fig:ch} Spectra of J0585$-$0809, J1432$-$4125, and J2005$-$3057 surrounding the Eu line at 4129\,{\AA} and the CH bandhead at 4313\,{\AA}. Black points represent the observed spectrum, while the solid colored lines represent the best-fit synthetic spectra. Dotted lines represent synthetic spectra within $\pm 0.3$\,dex of our adopted abundances.}
\end{center}
\end{figure*}

We then observed J0858$-$0809, J1432$-$4125, and J2005$-$3057 using the Magellan-Clay telescope and the MIKE spectrograph \citep{Bernstein03} at Las Campanas Observatory, on 2016 April 14 and 15. J0858$-$0809 was re-observed on 2016 April 17. For each star, we obtained a high-resolution spectrum with nominal resolving power of $R \sim 35,000$ in the blue and $R \sim 28,000$ in the red wavelength regime, using a $0\farcs7$ slit. The spectra cover $\sim$3350\,{\AA} to $\sim$9000\,{\AA}, with the blue and red CCDs overlapping at around $\sim5000$\,{\AA}.
Data reductions were completed using the MIKE Carnegie Python pipeline \citep{Kelson03}. To combine the data for J0585$-$0809 from both nights, we first reduced the data from each night separately. The reduced spectra were  co-added after shifting the spectrum from the second night of observations into the rest frame of the spectrum from the first night. We show two representative portions of the final spectra in Figure~\ref{fig:ch}, around the Eu line at 4129\,{\AA} and the CH bandhead at 4313\,{\AA}.  Additional details regarding our observations, including signal-to-noise ($S/N$) and heliocentric velocities, are listed in Table \ref{table:observations}.  

We measured heliocentric radial velocities (v$_{\mathrm{helio}}$) by cross-correlating our spectra against a template spectrum of HD~140283. Our results are $+169.5\pm 0.9$ \kms, 
$-228.7 \pm 1.3$\,\kms,  and $-265.9\pm1.4$\,\kms \ for J0858$-$0809, J1432$-$4125, and J2005$-$3057, respectively. We derive  uncertainties from the standard deviation of v$_{\mathrm{helio}}$ measurements found using several different template spectra. In the case of J0858$-$0809, we average the radial velocities measured on both observation nights ($+168.1\pm1.2$\,\kms and $+170.8\pm1.3$\,\kms, respectively) and add their individual uncertainties in quadrature. 

Previous survey observations also report v$_{\mathrm{helio}}$ values for J0858$-$0809, J1432$-$4125, and J2005$-$3057. RAVE DR5 \citep{kunder17} reports a heliocentric radial velocity of $+168.6 \pm 1.4$\,\kms \ for J0858$-$0809 from 2008 April 12, and  $-264.8 \pm 0.9$\,\kms \ (2007 September 6) and $-264.5 \pm 0.9$ (2008 May 23) for J2005$-$3057.
For completeness, we note that new Gaia DR2 measurements taken between 2014 July 25 and 2016 May 23 find v$_{\mathrm{helio}}=170.1\pm0.4$\,\kms and $-230.1\pm0.8$\,\kms for J0858$-$0809 and J1432$-$4125, respectively \citep{gaiab, gaiaa}. 

Our adopted heliocentric radial velocities for all three stars are consistent with previous measurements from RAVE DR5 and Gaia DR2 within one standard deviation, strongly suggesting that they are all single. 
This is in line with the vast majority of metal-poor $r$-process-enhanced stars, $\sim 82\%$ of which exhibit no radial velocity variations arising from a binary companion \citep{Hansen15}.

\subsection{Line Measurements}  
We performed a standard abundance analysis for our stars 
as described in \citet{frebel13a}. For our analysis, we used the latest version of the \texttt{MOOG} code\footnote{\href{url}{https://github.com/alexji/moog17scat} where Rayleigh scattering \citep{Sneden73, Sobeck11} is accounted for.} Our software employs a 1D plane-parallel model atmosphere with $\alpha$-enhancement \citep{Castelli04} and assumes local thermal equilibrium (LTE).  All line measurements, stellar parameters, and abundance measurements were made using custom \texttt{SMH} software, first described in \citet{Casey14}.

Iron equivalent widths were derived using a line list compiled with data from \citet{obrian91}, \citet{kurucz_lines}, \citet{mel09}, \citet{den14}, and \citet{ruf14}. Neutron-capture line lists used data from \citet{Hill02,hill17}. We used synthesis line lists provided by Chris Sneden, which are based on atomic data from \citet{Sneden09, sneden14, Sneden16}, and supplemented with data from \citet{kurucz_lines}. The CH synthesis line lists were taken from \citet{Masseron14}. \xfe \ values were calculated using solar abundances from \cite{Asplund09}. 

We measured line equivalent widths in \texttt{SMH} by performing $\chi^{2}$ minimized Gaussian fits of each observed line profile. We modeled the local continuum by masking absorption lines near the line of interest. In the case that a line was heavily blended or had hyperfine structure features, we performed spectrum synthesis using $\chi^2$ minimization to obtain the best fit. We used \texttt{SMH} synthesis tools to fit the line of interest 
and all surrounding lines within the local wavelength region using already measured abundances. For lines too small to be detected, we obtain a $3\sigma$ upper limit on the abundance. 
Our full equivalent width and synthesis measurements for J0858$-$0809, J1432$-$4125, and J2005$-$3057 are listed in Table~\ref{table:data}. The resulting abundances are discussed in greater detail in Section \ref{section:abundances}. 

\begin{figure}[!ht] 
\begin{center}
  \includegraphics[clip=false,width=0.40\textwidth]{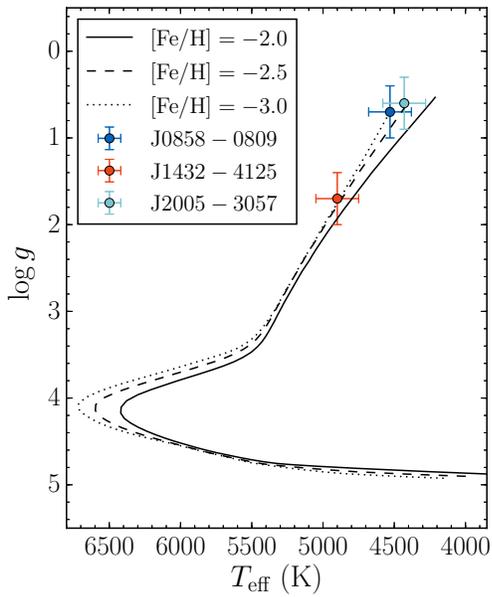}  
   \caption{\label{fig:isochrone} A 12 Gyr isochrone \citep{Kim02} displaying the stellar parameters for all three stars.  J0858$-$0809, J1432$-$4125, and J2005$-$3057 have metallicities of  $-3.16$, $-2.97$, and $-3.03$, respectively. }
\end{center}
\end{figure}

\section{Stellar parameters}\label{params}

Stellar parameters for J1432$-$4125, J0858$-$0809, and J2005$-$3057 were determined spectroscopically from Fe\,I and II lines using the procedure detailed in \citet{frebel13a}. We obtained all of our Fe line 
measurements through equivalent width analysis. In total, we measured 291 Fe\,I lines and 22 Fe\,II lines for J1432$-$4125, 283 Fe\,I lines and 27 Fe\,II lines for J0858$-$0809, and 251 Fe\,I lines and 29 Fe\,II lines for J2005$-$3057. 

We used an iterative method to determine the effective temperature \Teff, the surface gravity \logg, the metallicity \feh, and the microturbulence \vmic. 
Following first estimates based on the medium-resolution results, we fixed the temperature by forcing zero trend in a linear regression of Fe\,I line abundances and excitation potential. We then adjusted \vmic \ to achieve zero trend between the Fe\,I line abundances and reduced equivalent width. We obtained $\log g$ and \feh{ } by fine-tuning our parameters until ionization balance between Fe\,I and Fe\,II was achieved and the model atmosphere metallicity was consistent with the metallicity of our Fe\,I lines. We iteratively adjusted these parameters until total convergence was reached. Finally, we applied temperature corrections described in \citet{frebel13a} to obtain our adopted stellar parameters. Both corrected and uncorrected values based on our high-resolution spectra can be found in Table \ref{table:parameters}. We also provide results from the medium-resolution follow-up spectra, for comparison. We adopt the corrected stellar parameters as our final values.

We adopt typical systematic errors in the stellar parameters as obtained from a spectroscopic analysis \citep{frebel13a,Ji16a}. 
 We take $\sigma_{T_{\rm eff}} = 150$\,K, $\sigma_{\log g} = 0.30$\,dex, and $\sigma_{\textrm{v}_\textrm{micr}} = 0.30$\,\kms. Statistical contributions to $\sigma_{T_{\rm eff}}$, $\sigma_{\log g}$, and $\sigma_{\textrm{v}_\textrm{micr}}$ are negligible in comparison, due to the brightness of our stars ($V \sim 11$) and the large number of Fe\,I lines measured ($251$-$291$). The uncertainty in \feh \ for each star is derived from the standard deviation of Fe\,I line abundances, around $\sim0.15$ for each star.  Figure~\ref{fig:isochrone} displays our adopted LTE stellar parameters on a 12 Gyr isochrone that includes predictions for tracks with $\feh{ } = -2.0$, $\feh{ } = -2.5$, and $\feh{ } = -3.0$, derived from \cite{Kim02}. Our stellar parameters agree very well with the most metal-poor isochrone. Our results show that J0858$-$0809 and J2005$-$3057 are located at the tip of the red giant branch, whereas J1432$-$4125 is a warmer giant.

\begin{deluxetable}{lcccccccccc}[h!]
\tablecaption{Stellar Parameters \label{table:parameters}}
\tablehead{
\colhead{Star} &
\colhead{\Teff} &
\colhead{\logg} & \colhead{\vmic}&  \colhead{\feh} \\ 
\colhead{} &
\colhead{[K]} &
\colhead{[cgs]} & \colhead{[\kms]}&  \colhead{}} 
\startdata 
\multicolumn{5}{c}{LTE Parameters (corrected)}\\
\hline 
J0858$-$0809 &   $4530$&     $0.70$  &    $2.25$ &  $-3.16$\\
J1432$-$4125 & $4900$& $1.70$  &    $1.60$  & $-2.97$\\
J2005$-$3057 &   $4430$&    $0.60$ &   $2.30$  &  $-3.03$\\ \\
\multicolumn{5}{c}{LTE Parameters (uncorrected)}\\
\hline 
J0858$-$0809 &   $4290$&     $0.00$  &    $2.15$ &  $-3.36$\\
J1432$-$4125 & $4705$& $1.20$  &  $1.55$  & $-3.15$\\
J2005$-$3057 &   $4180$&    $0.00$ &   $2.15$  &  $-3.25$\\ \\
\hline
\multicolumn{5}{c}{NLTE Parameters} \\
\hline 
J0858$-$0809 &  4400 & 1.20 & 2.10 & $-2.81$     \\
J1432$-$4125 & 4850 &1.90 &1.50&$-2.85$  \\
J2005$-$3057 & 4300 &0.90 & $2.20$& $-2.85$   \\ \\
\hline 
\multicolumn{5}{c}{Parameters from Medium-Resolution Spectrum} \\
\hline
J0858$-$0809 &  $4770$ & $0.97$ & \nodata  & $-2.96$  \\
J1432$-$4125 & $5334$ & $2.64$ & \nodata & $-2.91$  \\
J2005$-$3057 &  $4599$ & $1.67$  & \nodata  & $-3.05$  \\
\enddata 
\tablecomments{Corrected LTE parameters are adopted for analysis.}
\end{deluxetable}

We also determined stellar parameters taking into account  non-LTE (NLTE) effects. Departures from LTE are more significant in metal-poor stars. Metal-poor stars have lower electron densities and fewer atomic collisions in their atmospheres, and are effectively radiatively dominated. This can cause line formation to deviate from LTE conditions. Deviations are especially common for minority species, such as Fe\,I. 

To obtain NLTE stellar parameters, we determined NLTE abundances from Fe\,I and Fe\,II lines. Starting from the LTE stellar parameters, we changed each parameter iteratively until excitation and ionization equilibrium were attained in NLTE between abundances of Fe\,I and Fe\,II lines, following the procedure outlined in \citet{Ezzeddine2017}. A comprehensive Fe atom was used \citep{Ezzeddine2016} with up-to-date atomic data, especially for hydrogen collisions from \citet{Barklem2018}. Our NLTE stellar parameters are also listed in Table \ref{table:parameters}. They are somewhat cooler but slightly more metal-rich than the corrected LTE values.

The results for J0858$-$0809, J1432$-$4125, and J2005$-$3057 are  consistent with the NLTE metallicities predicted by \citet{ezzeddine16} using \[\Delta [\mathrm{Fe/H}]= -0.14[\mathrm{Fe/H}]_{\mathrm{LTE}}-0.15.\] 
We find the difference in predicted and measured $\Delta[\mathrm{Fe/H}]$ values to be
$\Delta[\mathrm{Fe/H}]_{\rm predicted} - \Delta [\mathrm{Fe/H}]_{\rm measured} = -0.06$, $+0.15$ and $+0.09$\,dex for J0858$-$0809, J1432$-$4125, and J2005$-$3057, respectively. These differences are reasonable with respect to our reported uncertainties in [Fe/H], which are $0.15$, $0.13$, and $0.15$\,dex, respectively.  J0858$-$0809 and J2005$-$3057 (\Teff $= 4530$ and $4430$\,K) both have temperature corrections of $+130$\,K, whereas J1432$-$4125 has a smaller correction of $+50$\,K. These values are well within our adopted uncertainty of $\sigma_{T_{\rm eff}}= 150$\,K. They are also within or near the uncertainties reported in \citet{ezzeddine16}, which estimates  $\sigma_{T_{\rm eff}}= 112$\,K, $\sigma_{\log g}=0.45$\,dex, and $\sigma_{\mathrm{v}_\mathrm{micr}}=0.40$ \kms for metal-poor stars with detectable Fe\,II lines.

Our NLTE stellar parameters likely provide a more accurate description of the nature of our stars. However, because most literature data assumes LTE, for the remainder of this paper we will use line abundances calculated assuming LTE.

\begin{deluxetable*}{lrrrrrrrrrrrrrrrrrr}[h!]
\tabletypesize{\tiny}
\tablewidth{0pc}
\tablecaption{Magellan/MIKE Chemical Abundances}
\tablehead{\colhead{}&
\multicolumn{5}{c}{J0858$-$0809}&\colhead{}&\multicolumn{5}{c}{J1432$-$4125}&\colhead{}&\multicolumn{5}{c}{J2005$-$3057}&\colhead{}\\
\cline{2-6} \cline{8-12} \cline{14-18}\\
\colhead{Element} &
\colhead{$\lg\epsilon (\mbox{X})$} &
\colhead{\xh} &  \colhead{\xfe}& \colhead{N} & \colhead{$\sigma$} & \colhead{} &\colhead{$\lg\epsilon (\mbox{X})$} &
\colhead{\xh} &  \colhead{\xfe}& \colhead{N} & \colhead{$\sigma$} &  \colhead{} &
\colhead{$\lg\epsilon (\mbox{X})$} &
\colhead{\xh} &  \colhead{\xfe}& \colhead{N} & \colhead{$\sigma$} & \colhead{}} 
\startdata
C\,&$5.28$&$-3.15$&$0.01$&$2$&$0.05$&&$5.68$&$-2.75$&$0.22$&$2$&$0.05$&&$5.02$&$-3.41$&$-0.37$&$2$&$0.05$\\ 
C (corr)&\nodata&\nodata&$0.76$&\nodata&\nodata&&\nodata&\nodata&$0.43$&\nodata&\nodata&&\nodata&\nodata&$0.38$&\nodata&\nodata\\ 
O\,&$6.44$&$-2.25$&$0.91$&$1$&$0.20$&&\nodata&\nodata&\nodata&\nodata&\nodata&&$6.55$&$-2.13$&$0.90$&$1$&$0.20$\\ 
Na\,I&$3.45$&$-2.79$&$0.36$&$2$&$0.05$&&$3.62$&$-2.62$&$0.34$&$2$&$0.05$&&$3.74$&$-2.50$&$0.54$&$2$&$0.05$\\ 
Na\,I (NLTE) &\nodata&\nodata&$-0.04$&\nodata&\nodata&&\nodata&\nodata&$-0.01$&\nodata&\nodata&&\nodata&\nodata&$0.14$&\nodata&\nodata\\
Mg\,I&$4.96$&$-2.64$&$0.52$&$12$&$0.11$&&$5.11$&$-2.49$&$0.49$&$11$&$0.14$&&$5.07$&$-2.53$&$0.50$&$11$&$0.14$\\ 
Al\,I&$2.72$&$-3.73$&$-0.57$&$2$&$0.13$&&$2.78$&$-3.67$&$-0.71$&$2$&$0.08$&&$2.87$&$-3.58$&$-0.55$&$2$&$0.19$\\ 
Si\,I&$4.89$&$-2.62$&$0.54$&$2$&$0.13$&&$5.19$&$-2.32$&$0.64$&$2$&$0.19$&&$4.95$&$-2.56$&$0.47$&$2$&$0.05$\\ 
Ca\,I&$3.59$&$-2.75$&$0.41$&$24$&$0.09$&&$3.84$&$-2.50$&$0.47$&$26$&$0.09$&&$3.66$&$-2.68$&$0.36$&$22$&$0.09$\\ 
Sc\,II&$0.06$&$-3.09$&$0.07$&$13$&$0.12$&&$0.25$&$-2.90$&$0.07$&$10$&$0.10$&&$0.11$&$-3.04$&$-0.01$&$10$&$0.11$\\ 
Ti\,I&$1.99$&$-2.96$&$0.19$&$29$&$0.08$&&$2.29$&$-2.66$&$0.31$&$24$&$0.07$&&$2.06$&$-2.89$&$0.15$&$28$&$0.11$\\ 
Ti\,II&$2.04$&$-2.91$&$0.25$&$48$&$0.10$&&$2.33$&$-2.62$&$0.35$&$50$&$0.09$&&$2.19$&$-2.76$&$0.27$&$51$&$0.12$\\ 
V\,II&$0.89$&$-3.04$&$0.11$&$5$&$0.09$&&$1.05$&$-2.88$&$0.09$&$4$&$0.07$&&$0.86$&$-3.07$&$-0.03$&$4$&$0.12$\\ 
Cr\,I&$2.21$&$-3.43$&$-0.27$&$20$&$0.15$&&$2.45$&$-3.19$&$-0.22$&$18$&$0.10$&&$2.33$&$-3.31$&$-0.28$&$14$&$0.05$\\ 
Mn\,I&$1.64$&$-3.79$&$-0.64$&$8$&$0.19$&&$1.89$&$-3.54$&$-0.58$&$7$&$0.15$&&$1.70$&$-3.73$&$-0.70$&$7$&$0.15$\\ 
Fe\,I&$4.34$&$-3.16$&$0.00$&$283$&$0.15$&&$4.53$&$-2.97$&$0.00$&$291$&$0.13$&&$4.47$&$-3.03$&$0.00$&$251$&$0.15$\\ 
Fe\,II&$4.33$&$-3.17$&$-0.02$&$27$&$0.06$&&$4.53$&$-2.96$&$0.00$&$22$&$0.08$&&$4.47$&$-3.03$&$0.01$&$29$&$0.08$\\ 
Co\,I&$1.86$&$-3.13$&$0.03$&$9$&$0.17$&&$2.17$&$-2.82$&$0.15$&$6$&$0.10$&&$1.85$&$-3.14$&$-0.11$&$8$&$0.13$\\ 
Ni\,I&$3.10$&$-3.12$&$0.04$&$21$&$0.11$&&$3.27$&$-2.95$&$0.02$&$19$&$0.10$&&$3.12$&$-3.10$&$-0.07$&$19$&$0.12$\\ 
Zn\,I&$1.68$&$-2.88$&$0.28$&$2$&$0.06$&&$1.89$&$-2.67$&$0.29$&$2$&$0.10$&&$1.71$&$-2.85$&$0.18$&$2$&$0.05$\\ 
Sr\,II&$-0.47$&$-3.34$&$-0.18$&$2$&$0.05$&&$0.18$&$-2.69$&$0.28$&$2$&$0.05$&&$-0.17$&$-3.04$&$0.00$&$2$&$0.08$\\ 
Y\,II&$-1.31$&$-3.52$&$-0.36$&$11$&$0.09$&&$-0.60$&$-2.81$&$0.16$&$10$&$0.05$&&$-0.93$&$-3.14$&$-0.11$&$8$&$0.05$\\ 
Zr\,II&$-0.55$&$-3.13$&$0.03$&$5$&$0.12$&&$0.13$&$-2.45$&$0.52$&$7$&$0.11$&&$-0.20$&$-2.78$&$0.25$&$5$&$0.10$\\
Mo\,I&\nodata&\nodata&\nodata&\nodata&\nodata&&$-0.25$&$-2.13$&$0.84$&$1$&$0.20$&&$<-0.43$&$<-2.31$&$<0.72$&$1$&$0.20$\\ 
Ru\,I&$<-0.72$&$<-2.47$&$<0.69$&$1$&$0.30$&&$-0.02$&$-1.77$&$1.20$&$3$&$0.09$&&$-0.42$&$-2.17$&$0.86$&$1$&$0.30$\\ 
Rh\,I&\nodata&\nodata&\nodata&\nodata&\nodata&&$<0.17$&$<-0.74$&$<2.23$&$1$&$0.20$&&\nodata&\nodata&\nodata&\nodata&\nodata\\ 
Pd\,I&\nodata&\nodata&\nodata&\nodata&\nodata&&$-0.12$&$-1.69$&$1.28$&$1$&$0.20$&&\nodata&\nodata&\nodata&\nodata&\nodata\\ 
Ba\,II&$-1.32$&$-3.50$&$-0.34$&$5$&$0.09$&&$0.03$&$-2.15$&$0.82$&$5$&$0.13$&&$-0.67$&$-2.85$&$0.19$&$4$&$0.05$\\ 
La\,II&$-2.22$&$-3.32$&$-0.16$&$6$&$0.05$&&$-0.86$&$-1.96$&$1.01$&$17$&$0.05$&&$-1.39$&$-2.49$&$0.55$&$15$&$0.11$\\ 
Ce\,II&$-1.89$&$-3.47$&$-0.31$&$7$&$0.12$&&$-0.53$&$-2.11$&$0.86$&$20$&$0.08$&&$-1.14$&$-2.72$&$0.31$&$25$&$0.11$\\ 
Pr\,II&$-2.28$&$-3.00$&$0.16$&$3$&$0.08$&&$-0.98$&$-1.70$&$1.27$&$10$&$0.06$&&$-1.58$&$-2.30$&$0.73$&$9$&$0.07$\\ 
Nd\,II&$-1.63$&$-3.05$&$0.11$&$20$&$0.13$&&$-0.34$&$-1.76$&$1.21$&$54$&$0.08$&&$-0.93$&$-2.35$&$0.68$&$48$&$0.11$\\ 
Sm\,II&$-1.88$&$-2.84$&$0.32$&$1$&$0.30$&&$-0.67$&$-1.63$&$1.33$&$12$&$0.05$&&$-1.20$&$-2.16$&$0.87$&$9$&$0.08$\\ 
Eu\,II&$-2.41$&$-2.93$&$0.27$&$4$&$0.05$&&$-1.01$&$-1.53$&$1.44$&$5$&$0.09$&&$-1.57$&$-2.09$&$0.94$&$5$&$0.07$\\ 
Gd\,II&$-1.88$&$-2.95$&$0.2$&$3$&$0.07$&&$-0.51$&$-1.58$&$1.38$&$10$&$0.09$&&$-1.11$&$-2.18$&$0.85$&$6$&$0.08$\\ 
Tb\,II&\nodata&\nodata&\nodata&\nodata&\nodata&&$-1.22$&$-1.52$&$1.44$&$2$&$0.07$&&\nodata&\nodata&\nodata&\nodata&\nodata\\ 
Dy\,II&$-1.83$&$-2.93$&$0.23$&$3$&$0.08$&&$-0.39$&$-1.49$&$1.47$&$5$&$0.06$&&$-0.92$&$-2.02$&$1.18$&$4$&$0.10$\\ 
Ho\,II&$-2.38$&$-2.86$&$0.30$&$1$&$0.20$&&$-1.18$&$-1.66$&$1.31$&$10$&$0.07$&&$-1.71$&$-2.19$&$0.85$&$4$&$0.12$\\ 
Er\,II&$-2.08$&$-3.00$&$0.16$&$4$&$0.22$&&$-0.66$&$-1.58$&$1.39$&$10$&$0.09$&&$-1.19$&$-2.11$&$0.92$&$4$&$0.17$\\
Tm\,II&\nodata&\nodata&\nodata&\nodata&\nodata&&$-1.45$&$-1.55$&$1.41$&$5$&$0.14$&&$-2.15$&$-2.25$&$0.78$&$3$&$0.05$\\ 
Yb\,II&$-2.26$&$-3.10$&$0.06$&$1$&$0.10$&&$-1.41$&$-2.25$&$0.78$&$1$&$0.10$&&$-0.79$&$-1.63$&$1.43$&1&0.10\\ 
Hf\,II&\nodata&\nodata&\nodata&\nodata&\nodata&&$-0.84$&$-1.69$&$1.27$&$4$&$0.26$&&$-1.70$&$-2.55$&$0.48$&1&0.30\\ 
Os\,I&$<-1.13$&$<-2.53$&$<0.63$&$1$&$0.30$&&$0.00$&$-1.40$&$1.57$&$2$&$0.05$&&$-0.64$&$-2.04$&$1.00$&$2$&$0.19$\\ 
Ir\,I&$<-1.00$&$<-2.38$&$<0.78$&$1$&$0.30$&&$-0.12$&$-1.50$&$1.46$&$2$&$0.05$&&$-0.46$&$-1.83$&$1.20$&$1$&$0.20$\\ 
Th\,II&$-3.07:$
&$-3.09$&0.06&1&0.30&&$-1.47$&$-1.49$&$1.48$&$1$&$0.10$&&$-2.18$&$-2.20$&$0.84$&$1$&$0.10$\\
\enddata
\tablecomments{The abundance uncertainty $\sigma$ is derived from the standard deviation of individual line abundances. We calculate appropriate uncertainties for small samples for elements with 2-5 lines. For elements with one line, we adopt an uncertainty between 0.1$-$0.3\,dex based on the quality of the measurement. }
\label{table:abund}
\end{deluxetable*} 

\section{Chemical Abundances}\label{section:abundances}
We obtained abundance measurements for J0858$-$0809, J1432$-$4125, and
J2005$-$3057 using a mixture of spectrum synthesis and equivalent
width analysis. We estimate abundance uncertainties based on the
spread in line abundances and the data and fit quality. The standard
error in line abundances for most elements is small ($\sim
0.01$\,dex), as it does not fully account for uncertainties due to
data quality. Realistically, precision better than 0.05\,dex is
improbable, due to $S/N$ and associated continuum placement
difficulties. Thus, we derive the statistical uncertainty in
abundance, $\sigma$, for each element from the standard deviation of
the corresponding individual line abundances.  Other systematic
uncertainties, e.g., due to NLTE effects, 1D stellar model
atmospheres, and gf-values are not explicitly considered. For
  elements with one line, we adopt an uncertainty between 0.1 and
  0.3\, dex, depending on the data and fit quality.  For elements with
  $2 \leq \mathrm{N} \leq 5$ lines, we use small sample
    statistics to estimate an unbiased standard deviation of the line
    abundances. Following \citet{Keeping62}, we multiply the range of
    values covered by our line abundances with the k-factor calculated
    from small samples. This ensures that we are not underestimating
    uncertainties that would ordinarily be obtained from assuming $N$
    to be statistically meaningful.  We adopt minimum uncertainties
  of 0.05\,dex for elements with unreasonably small calculated
  uncertainties.

Table \ref{sys_errors} enumerates the systematic errors in our chemical abundances resulting from uncertainties in our model atmosphere parameters. We obtain these systematic errors by varying the stellar parameters by their uncertainties ($\Delta$ \Teff \ = 150 K, $\Delta$ \logg \ = 0.30\,dex, $\Delta$ \vmic \ = 0.30 \kms, $\Delta$\feh $\ \sim 0.15$\,dex, depending on the star), and finding the resulting change in abundances.

We now describe in detail our abundance measurements, which are summarized in Table \ref{table:abund} and fully detailed in Table \ref{table:data}. Figure~\ref{fig:bigplot} displays the lighter element abundances of our stars compared with previously observed Milky Way halo stars from \citet{yong13_II}. Figure~\ref{fig:rproc_pattern} displays our \ncap abundances on a scaled solar \rproc pattern from \citet{Burris00}.

\begin{deluxetable*}{lccccccccccccccc}
\tabletypesize{\tiny}
\tablewidth{0pc}
\tablecaption{Systematic errors}
\tablehead{\colhead{}&
\multicolumn{4}{c}{J0858$-$0809}&\colhead{}&\multicolumn{4}{c}{J1432$-$4125} &\colhead{} & \multicolumn{4}{c}{J2005$-$3057} & \colhead{}\\
\cline{2-5} \cline{7-10} \cline{12-15}\\
\colhead{Element} &
\colhead{$\Delta T_\textrm{eff}$} &
\colhead{$\Delta\log(\mbox{g})$} &  \colhead{$\Delta\textrm{$v$}_\textrm{micr}$}& \colhead{Root Mean} & \colhead{} &
\colhead{$\Delta T_\textrm{eff}$} &
\colhead{$\Delta\log(\mbox{g})$} &  \colhead{$\Delta\textrm{$v$}_\textrm{micr}$}& \colhead{Root Mean} & \colhead{} &
\colhead{$\Delta T_\textrm{eff}$} &
\colhead{$\Delta\log(\mbox{g})$} &  \colhead{$\Delta\textrm{$v$}_\textrm{micr}$}& \colhead{Root Mean} & \colhead{}\\
\colhead{} &
\colhead{+150\,K} &
\colhead{+0.30\,dex} &  \colhead{+0.30\,dex}& \colhead{Square} & \colhead{} &
\colhead{+150\,K} &
\colhead{+0.30\,dex} &  \colhead{+0.30\,dex}& \colhead{Square} & \colhead{} &
\colhead{+150\,K} &
\colhead{+0.30\,dex} &  \colhead{+0.30\,dex}& \colhead{Square} & \colhead{}
} 
\startdata
C\,&+0.40&$-0.12$&+0.01&0.42&&+0.32&$-$0.12&+0.02&0.34&&+0.35&$-$0.09&+0.01&0.36\\
O\,I&+0.13&+0.09&+0.00&0.16&&\nodata&\nodata&\nodata&\nodata&&+0.10&+0.09&$-$0.01&0.13\\
Na\,I&+0.16&$-0.07$&$-$0.18&0.25&&+0.19&$-$0.04&$-$0.17&0.26&&+0.26&$-$0.07&$-$0.18&0.32\\ 
Mg\,I&+0.14&$-0.07$&$-$0.06&0.17&&+0.15&$-$0.07&$-$0.05&0.17&&+0.17&$-$0.08&$-$0.07&0.20\\ 
Ca\,I&+0.10&$-$0.04&$-$0.02&0.11&&+0.11&$-$0.02&$-$0.03&0.12&&+0.13&$-$0.04&$-$0.03&0.14\\ 
Ti\,I&+0.20&$-$0.05&$-$0.02&0.21&&+0.18&$-$0.02&$-$0.02&0.18&&+0.22&$-$0.05&$-$0.03&0.23\\ 
Ti\,II&+0.07&+0.08&$-$0.06&0.12&&+0.07&+0.08&$-$0.07&0.13&&+0.05&+0.08&$-$0.07&0.12\\
Cr\,I&+0.19&$-$0.05&$-$0.05&0.20&&+0.18&$-$0.03&$-$0.06&0.19&&+0.21&$-$0.04&$-$0.01&0.21\\
Fe\,I&+0.17&$-$0.04&$-$0.05&0.18&&+0.17&$-$0.02&$-$0.07&0.18&&+0.20&$-$0.04&$-$0.05&0.21\\ 
Fe\,II&+0.01&+0.09&$-$0.02&0.09&&+0.01&+0.09&$-$0.04&0.10&&$-$0.01&+0.10&$-$0.03&0.10\\
Ni\,I&+0.16&$-$0.04&$-$0.05&0.17&&+0.18&$-$0.04&$-$0.13&0.23&&+0.20&$-$0.04&$-$0.06&0.21\\
Zn\,I&+0.07&+0.04&$-$0.01&0.08&&+0.07&+0.04&$-$0.01&0.08&&+0.06&+0.05&+0.00&0.08\\
Sr\,II&+0.12&+0.05&$-$0.30&0.30&&+0.10&+0.03&$-$0.32&0.34&&+0.15&+0.06&$-$0.26&0.31\\ 
Ba\,II&+0.13&+0.07&$-$0.08&0.22&&+0.12&+0.08&$-$0.21&0.25&&+0.14&+0.07&$-$0.17&0.23\\
Ce\,II&$-$0.21&+0.16&+0.06&0.27&&$-$0.10&+0.09&$-$0.01&0.13&&+0.11&+0.10&+0.00&0.15\\
Nd\,II&+0.15&+0.09&+0.00&0.17&&+0.12&+0.09&$-$0.01&0.15&&+0.12&+0.09&$-$0.01&0.15\\
Eu\,II&+0.15&+0.06&+0.02&0.18&&+0.11&+0.08&$-$0.01&0.14&&+0.09&+0.07&$-$0.02&0.12\\ 
Er\,II&+0.16&+0.10&$-$0.01&0.19&&+0.08&+0.07&$-$0.05&0.21&&+0.11&+0.09&$-$0.06&0.15\\ 
Os\,I&\nodata&\nodata&\nodata&\nodata&&+0.25&+0.10&+0.05&0.26&&+0.37&+0.05&+0.06&0.38\\ 
Th\,II&\nodata&\nodata&\nodata&\nodata&&+0.15&+0.11&+0.01&0.19&&+0.09&+0.13&+0.03&0.16\\
\enddata
\end{deluxetable*} \label{sys_errors}

\subsection{Light Elements}
\begin{figure*}[t!] 
\begin{center}
  \includegraphics[width=.9\textwidth]{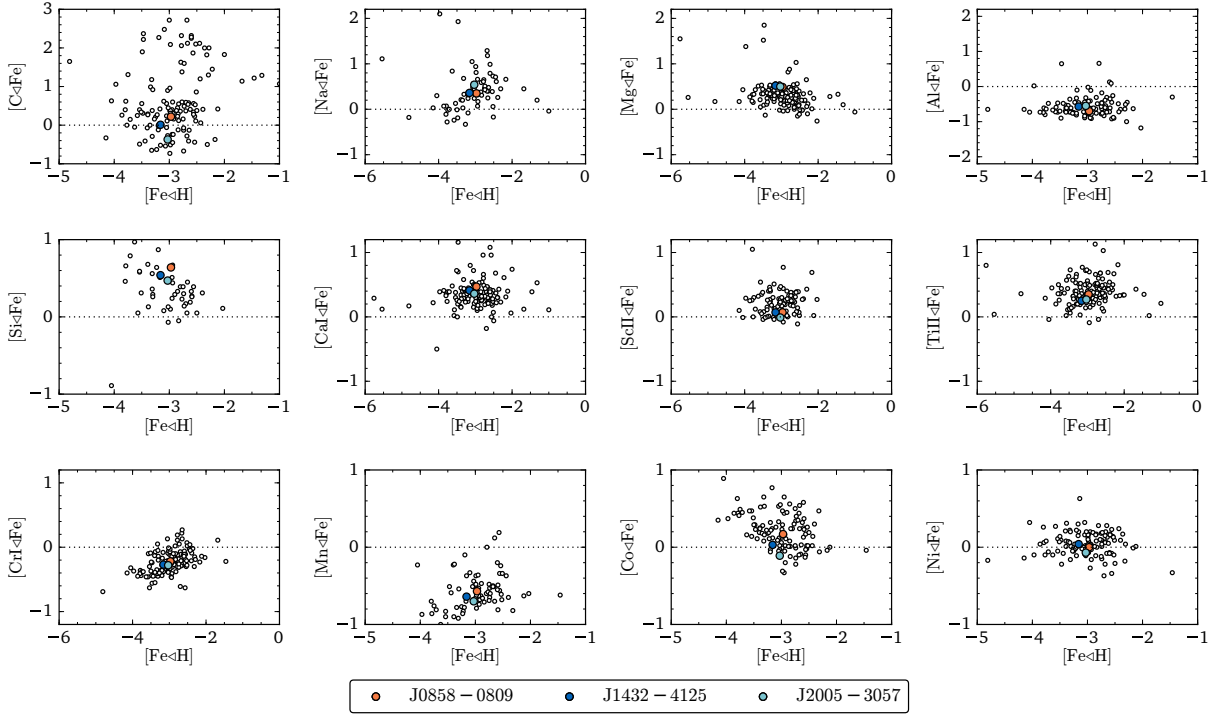} 
   \caption{Light-element abundances of J0858$-$0809, J1432$-$4125, and J2005$-$3057 overlaid with literature data for other metal-poor stars from \citet{yong13_II}. Data for C and Na is not corrected for evolutionary status or NLTE behavior.}
   \label{fig:bigplot}
\end{center}
\end{figure*}

We measured the carbon abundances of our stars by performing spectrum synthesis on the CH G-bandhead at 4313 \AA \ and the CH feature at 4323 \AA. Synthesis fits of the line at 4313\,\AA{ } are displayed in Figure \ref{fig:ch}. We determined the $^{12}$C/$^{13}$C ratio for each star by fitting the lines at 4217\,\AA{ }and 4225\,\AA, and checking for good agreement at 4019\,\AA{ }and 4302\,\AA. $\AB{C}{Fe}{ }$ values for all three stars ranged between $-0.37$ and $+0.22$\,dex. These relatively low carbon abundances are partially explained by the evolutionary status of our stars. All three stars are red giants, meaning that the CNO cycle has depleted their carbon abundances during evolution along the giant branch. 
Therefore, we apply abundance corrections using our LTE stellar parameters to account for the carbon depletion due to our stars' evolutionary statuses, following \citet{Placco14}. The corrected \AB{C}{Fe}{ } values, which are expected to represent the composition of the natal gas clouds, are +0.76 for J0858$-$0809, +0.43 for J1432$-$4125, and +0.38 for J2005$-$3057. NLTE stellar parameters yield similar  corrected [C/Fe] values of +0.65, +0.29, and +0.38 for J0858$-$0809, J1432$-$4125, J2005$-$3057, respectively. We adopt the LTE carbon corrections for consistency.

From these corrections, we determine that J0858$-$0809 ($\mbox{Eu/Fe}=0.23$, \feh $=-3.16$) is a carbon-enhanced metal-poor star (CEMP, \AB{C}{Fe}{ }$> +0.70$; \citealt{Aoki07}) that exhibits the $r$-process signature pattern. Hence, J0858$-$0809 adds to the small group of CEMP-$r$ stars.
Very few ($\sim10$) CEMP-$r$ stars have been observed so far (CS~22892-052 is the most well-known example), despite the fact that $\sim 43\%$ of observed non-CEMP-$r/s$ halo stars with \feh \ $\leq -3.0$ are carbon enhanced \citep{Placco14}. We speculate that more CEMP-$r$ stars are discoverable among $r$-process stars with low [Eu/Fe] values, as these are expected to be more common than more highly enriched $r$-process stars, and thus perhaps more representative of the typical metal-poor halo population. 

Light-element abundances, including O, Na, Mg, Al, Si, Ca, Sc, Ti, V, Cr, Mn, Co, and Ni, were obtained using a mixture of equivalent width measurements and spectrum synthesis.  We measured the Si line at 3905\,\AA{ } using spectrum synthesis in all three stars. For J0858$-$0809 and J2005$-$3057, we also measured the equivalent width of the Si I line at 4102\,\AA. The abundances of the $\alpha$ elements for all three stars are remarkably consistent. All three stars have $\AB{Mg}{Fe}{ } \sim +0.50$, $\AB{Si}{Fe}{ } \sim +0.55$, $\AB{Ca}{Fe}{ } \sim +0.40$, and $\AB{Ti}{Fe} \sim +0.30$. This level of $\alpha$-element enhancement ($[\alpha/\mathrm{Fe}] \sim +0.4$) is consistent with stars whose abundance enhancement originates primarily from core-collapse supernovae, rather than Type Ia supernovae. We measure aluminum from synthesis measurements of the Al~I line at 3944\,\AA{ } and equivalent width measurements of the Al I at 3961\,\AA. Sodium abundances are derived from the equivalent widths of the Na doublet at 5890\,\AA{ } and 5895\,\AA. We apply non-LTE corrections to the Na abundance of J1432$-$4125 using results from \citet{Lind11}. Standard non-LTE corrections of $-0.40$\,dex are applied to the Na abundances of J0858$-$0809 and J2005$-$3057 \citep{gehren2004_nlte}, as these stars are too cool for the corrections from \citet{Lind11} to apply. Our Na corrections are listed in Table \ref{table:abund}. Overall, all light-element abundances are in strong agreement with metal-poor stars analyzed by \cite{yong13_II}, as can be seen in Figure~\ref{fig:bigplot}.

\subsection{Neutron-Capture Elements}

We derive abundances for up to 25 neutron-capture elements, depending on the star. Measurements and $3\sigma$ upper limits for Sr, Y, Zr, Mo, Ru, Rh, Pd, Ba, La, Pr, Sm, Eu, Tb, Dy, Ho, Tm, Yb, Hf, Yb, Os, Ir, and Th were measured with spectrum synthesis, which accounts for hyperfine structure and blending of absorption features. Abundances based on equivalent widths were obtained for Ce, Nd, Gd, and Er.  For Ba and Eu measurements, we used \rproc isotope ratios  as given in \citet{Sneden08}. Abundance results and uncertainties are given in Table~\ref{table:abund}. The full set of line abundances and associated atomic data of all measured elements are presented in Table~\ref{table:data}, for reference.

\begin{figure*}[tbh!] 
  \includegraphics[clip=false,width=\textwidth,]{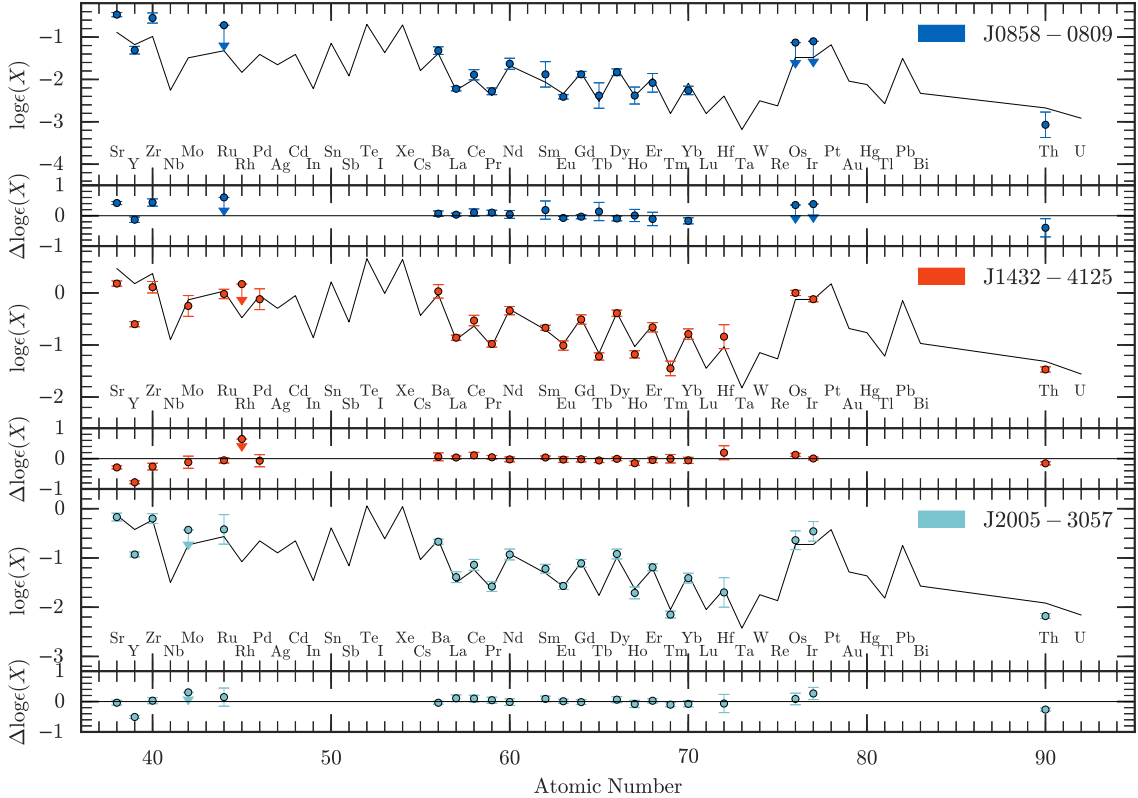}
   \caption{
 $R$-process elemental abundance patterns for J0858$-$0809, J1432$-$4125, and J2005$-$3057 overlaid with a scaled solar $r$-process pattern from \cite{Burris00}. Residuals are shown as well. } 
 \label{fig:rproc_pattern}
\end{figure*}

 Figure~\ref{fig:rproc_pattern} displays the \ncap element abundances of our three stars, overlaid with the scaled solar $r$-process patterns of \cite{Burris00}. The scaling was determined from a $\chi^2$ minimization on the square of the residual of \rproc elements with $56\le Z < 76$, weighted by the inverse abundance error, as in \cite{Ji16b}. Explicitly, the scaling factor is given by \[\underset{\epsilon \mathrm{\,offset}}{\mathrm{min}} \sum_{Z} \Big(\frac{\log \epsilon (Z_{\mathrm{\star}}) - \log \epsilon (Z_{\mathrm{\sun}}) + \epsilon_{\mathrm{offset}}}{\sigma_Z}\Big)^2,\] where $\sigma_Z$ is the abundance uncertainty of element $Z$. Overall, we find excellent agreement between the scaled solar pattern and the main \rproc abundances for element $Z\ge56$. Reduced $\chi^2$ values are $\chi^2/\nu = 0.78$, $0.74, $ and $0.57$ ($p = 0.34$, $0.27$, $0.13$) for J0858$-$0809, J1432$-$4125, and J2005$-$3057, where $\nu$ is the number of degrees of freedom ($\nu = 10$, $13$, $12$). We find mean residual standard deviations of $0.09$, $0.08$, $0.06$\,dex for J0858$-$0809, J1432$-$4125, and J2005$-$3057, respectively. These values, which indicate the spread from the scaled solar pattern, are on the order of typical statistical abundance uncertainties ($\sim0.05$-$0.10$\,dex). This provides yet more evidence for the universality of the main $r$-process pattern. 

In contrast, we find large deviations among light neutron-capture elements from the scaled \rproc solar pattern. The Sr residuals, which are representative of the light-element enhancement with respect to the scaled solar pattern, are $+0.42\pm0.05$, $-0.29\pm0.05$, and $-0.04\pm0.08$\,dex for J0858$-$0809, J1432$-$4125, J2005$-$3057, respectively. Here, we take the error to be the statistical uncertainty in the Sr abundance. A nearly identical trend is present in Zr residuals, which are $+0.44\pm0.12$, $-0.26\pm0.11$, and $+0.03\pm0.10$\,dex, respectively. Y residuals differ from Sr and Zr residuals as a result of the choice of scaled solar \rproc pattern \citep{Burris00}. Thus, we do not consider them in our analysis, though we note that the abundance differences $\log \epsilon (\mathrm{Sr})_\star-\log \epsilon (\mathrm{Y})_\star$ and $\log \epsilon (\mathrm{Zr})_\star-\log \epsilon (\mathrm{Y})_\star$ values are remarkably consistent in all three stars, ranging from 0.76-0.88\,dex and 0.73-0.76\,dex, respectively. This suggests that the same process produces Sr, Y, and Zr in a characteristic pattern. On the other hand, Sr and Zr residuals for J0858$-$0809 and J2005$-$3057 are statistically significant compared to the small variations due abundance uncertainties among heavier elements with $Z\ge56$.

\newpage
\section{Discussion and Conclusion}\label{discussion}

We have presented the full elemental abundance patterns for three new metal-poor $r$-process-enhanced red giant stars, J0858$-$0809 with $\mbox{[Eu/Fe]}=+0.23$, the $r$-I star J2005$-$3057 with $\mbox{[Eu/Fe]}=+0.94$, and the $r$-II star J1432$-$4125 with $\mbox{[Eu/Fe]}=+1.44$. All three stars exhibit remarkable agreement with the main component of the respectively scaled solar $r$-process patterns, i.e., for elements Ba and above. This universality has been seen again and again in $r$-process-enhanced stars, irrespective of their overall [Eu/Fe] values. 

Elements associated with the main $r$-process component are believed to be made in neutron star mergers. Nucleosynthesis calculations suggest that interactions between the  two inspiraling neutron stars (e.g., tidal ejecta) and dynamical ejecta during the merger itself provide a very large neutron-to-seed-ratio that enables production of elements up to and including U (e.g., \citealt{thiel17}). Given the abundant astrophysical evidence (e.g., \citealt{Hotokezaka15,Wallner15,Ji16a,drout17,kilpatrick17,shappee17}), it appears that the nucleosynthetic products of these two types of ejecta (possibly from the same site) must consistently yield what is observed as the universal main $r$-process component in the oldest stars. Alternatively, perhaps only one of the two types of ejecta is actually responsible for the observed abundance patterns. 

Neutron-capture elements lighter than Ba are believed to be made at least partially in a limited $r$-process \citep{frebel18}. Elements lighter than Cd (i.e., around the first $r$-process peak) are likely made exclusively in the limited $r$-process. The respective portion of the stellar abundance patterns indeed suggests no universality. In contrast, the origin of elements with $49\leq Z\leq 56$ may be attributed to the limited or main \rproc\citep{roederer12}. Unfortunately, only few elements are available and then they are only measurable in few stars. With more data hopefully coming in soon, better constraints on this element range can be obtained.
 
Core-collapse supernovae are obvious candidates for hosting the limited $r$-process as they are thought to provide only small yields of neutron-capture elements (especially the lighter neutron-capture elements). In addition, limited $r$-process events do not seem to correlate with a neutron star merger event that would have produced the heavy neutron-capture elements. 

This behavior is reflected in how the light elements, Sr and Zr, deviate from the scaled solar pattern (see Figure~\ref{fig:rproc_pattern});  J0858$-$0809 has positive residuals, J1432$-$4125 has negative residuals, and J2005$-$3057 has Sr and Zr abundances basically in agreement with the  solar pattern scaled to the heavier elements. (We note here that a portion of the disagreement of the Y abundances result from the choice of the solar $r$-process pattern, although overall, the deviations of Y follows in lockstep with Sr and Zr.) Hence, J0858$-$0809 may have formed in an environment that was enriched by a larger supernova-to-neutron star merger ratio than J1432$-$4125 or J2005$-$3057. 

Assuming that halo $r$-process-enhanced stars form in small, early galaxies similar to that of, e.g., Tucana\,III \citep{hansen17} that are later accreted by the Milky Way, this ratio would  reflect the naturally varying number of early supernovae in each system. A similar conclusion was suggested for the $r$-process stars recently found in Reticulum\,II \citep{Ji16b}. All their stars exhibit among the lowest known Sr, Y, Zr abundances measured in \rproc enhanced stars with respect to the Solar $r$-process pattern scaled to elements Ba and above. This could be understood if the low-mass system Reticulum\,II experienced a smaller number of supernovae compared to other, more massive systems such as Tucana\,III (it must have been more massive in the past since it is currently being tidally disrupted; \citealt{simon17}) whose $r$-process star shows a closer agreement to the overall scaled solar $r$-process.

Alternatively, variations in the $r$-process yields or potential sites might be able to explain the variations seen in the abundance data.  Examples include light neutron-capture elements being made in a limited $r$-process in the accretion disks around a merged pair of neutron stars or in the shock-heated ejecta that emerge during the merger. However, at least some supernovae can be expected to explode in a star-forming region prior to the formation of any subsequent stars. This implies that 
small amounts of light neutron-capture elements were in all likelihood provided unless supernovae are not producing any neutron-capture elements. Still, the combined yields of accretion disk/shock-heated ejecta and supernovae should be carefully considered to see if this could explain the observed light neutron-capture element abundance variations in $r$-process stars.

\begin{figure}
  \includegraphics[clip=false,width=.5\textwidth]{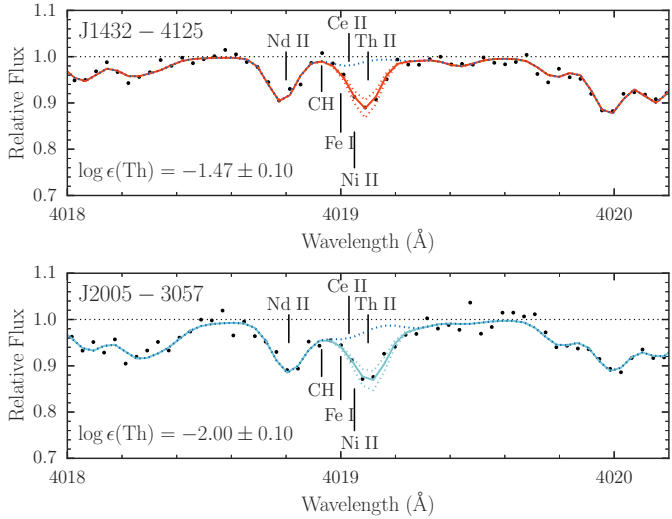} 
   \caption{\label{fig:thorium_plot} Spectra for J1432$-$4125 and J2005$-$3057 near the Th II line at 4019 \AA. Observed spectra are denoted in black, measured synthetic spectra are solid colored lines, and $\pm 0.10$\,dex synthetic spectra are dotted colored lines. We also show synthetic spectra with no Th contribution (dark blue dotted lines).}
\end{figure}

Another ``natural" deviation from the scaled solar $r$-process pattern stems from the radioactive decay of the long-lived radioactive \ncap elements such as thorium. By measuring the depletion of Th with respect to stable \rproc elements, stellar ages can be obtained. The Th\,II line at 4019\,{\AA} was detected in both J1432$-$4125 and J2005$-$3057, as seen in Figure~\ref{fig:thorium_plot}. We measured this line in J0858$-$0809 as well, but we do not use the resulting abundance for any age calculation. It has a large uncertainty (0.30\,dex) that mostly results from carbon enhancement and associated blending of the Th\,II line at 4019\,{\AA} with $^{13}$CH, Fe, Ni, and Ce. Fortunately, J1432$-$4125 and J2005$-$3057 are both carbon-poor with [C/Fe$] = +0.22$ and [C/Fe$] = -0.37$, respectively, so blending was minimal and the abundances could be measured to within $0.10$\,dex. 

To compute the age of J1432$-$4125 and J2005$-$3057, we compare measured $\log \epsilon (\mathrm{Th}/r)$ values to theoretical \rproc production ratios, which we denote as $\log \epsilon (\mathrm{Th}/r)_{\rm initial}$. Here $r$ represents a stable \rproc element above Ba.  We employ the formula \[\Delta t = 46.78[\log \epsilon {\text{(Th/r)}}_{\text{initial}} - \log \epsilon {(\text{Th/r})_{\textrm{now}}}]\] derived by  \cite{Cayreletal:2001} to estimate $\Delta t$, i.e., the time that has passed since the nucleosynthesis event that produced the main $r$-process elements. The half life of Th is contained in the leading constant. We adopt $\Delta t$ as the age of our star. We use production ratios from  \citet{schatz02}, which employs a  site-independent classical \rproc model with waiting point approximations. Production ratios from \citet{hill17}, based on high-entropy neutrino wind models from \citet{Farouqi10}, are also used for comparison. 

We propagate abundance uncertainties into our age uncertainty with the formula \[\sigma_{\Delta t}(\mathrm{X}) = 46.78 \sqrt{\sigma_{\log \epsilon (\mathrm{Th})}^2+\sigma_{\log \epsilon (\mathrm{X})}^2}. \] Here, $\sigma_{\log \epsilon (\mathrm{X})}$ represents the abundance uncertainty for element X, as discussed in Section~\ref{section:abundances}. A full list of abundance uncertainties are listed in Table~\ref{table:abund}. We take $\sigma_{[\mathrm{Th/Fe}]} = 0.10$\,dex. Note that we do not account for the systematic uncertainties in the initial production ratios or \rproc pattern, although they could be significant given that they remain poorly understood.

We calculate our final ages and uncertainties by averaging individual ages and uncertainties listed in Table~\ref{table:ages} for each set of production ratios. Note that we do not include ages from Hf, Os, and Ir abundances due to their large uncertainties of $\sim 0.30$\,dex. Neither do we include the Th/Sm initial production value of \citet{hill17} since they consistently give ages $\sim$10\,Gyr off from the average, suggesting an underlying systematic issue with this element ratio.

As can be seen, the individual ages vary significantly for each set of production ratios. In addition, the two sets of production ratios yield different results on average. Specifically, for J1432$-$4125, the \citet{hill17} ratios yield an age that is aligned with expectations, i.e., 12\,Gyr. For J2005$-$3057, the \citet{schatz02} production ratios yield the ``better" age of 10\,Gyr. This discrepancy highlights the need for additional  calculations of initial production ratios explicitly accounting for the astrophysical site, e.g., a neutron star merger. In the meantime, we adopt the above values as the best estimate of the ages of the two stars. Associated uncertainties are 6\,Gyr.

Hence, we take these ages as indicators of stellar age rather than precise values, given the large observational uncertainties as well as additional systematic uncertainties arising from the initial production ratios. For stars with strong $r$-process enhancement for which very high S/N data can be obtained, observational uncertainties can also be reduced (e.g., \citealt{Frebel07,placco17}). Regardless, our results here verify that J2005$-$3057 and J1432$-$4125 are indeed old stars, as suggested by their low metallicities. Additional high-resolution observations are underway to attempt a uranium measurement in at least one of these stars. A uranium measurement of an \rI star has yet to be made. 
 
We have confirmed that J0858$-$0809, J1432$-$4125, and J2005$-$3057 are extremely metal poor ([Fe/H] $\sim -3.0$) $r$-process-enhanced enhanced stars desprite their varying Eu enhancement. These stars were discovered as part of the $R$-Process Alliance, a new effort to uncover $r$-process-enhanced stars in Galactic halo to advance our understanding of the $r$-process and its astrophysical site. All three stars strongly follow the characteristic main \rproc pattern, though J0858$-$0809 and J1432$-$4125 demonstrate notable deviations in light \ncap element abundances with respect to the scaled solar pattern. This suggests that light and heavy \ncap elements may be produced by different \rproc sites. Future efforts by the $R$-Process Alliance should yield abundances for many more r-process-enhanced stars to further address these questions. This should also facilitate much-needed detailed comparison of $r$-process nucleosynthesis models for a variety of sites and conditions with ample observational data.

\begin{deluxetable}{lrrrrrll} 
\tabletypesize{\normalsize}
\tablecaption{Stellar ages derived from abundance ratios }
\tablehead{
\colhead{Th$/$r} & \colhead{PR\tablenotemark{i}} & \colhead{Age (Gyr)} & \colhead{PR\tablenotemark{ii}} & \colhead{Age (Gyr)} &\colhead{$\sigma_{\Delta t}$ (Gyr)}} 
\startdata  
\hline 
\multicolumn{6}{c}{J1432$-$4125} \\
\hline
Th/Ba &\nodata& \nodata & $-1.058$ & $20.68$&$7.67$\\
Th/La & $-0.60$ & $0.47$ & $-0.362$ & $11.60$& 5.23\\
Th/Ce & $-0.79$ & $7.02$ & $-0.724$ & $10.10$&$5.99$\\
Th/Pr & $-0.30$ & $8.89$ & $-0.313$ & $8.28$&$5.46$\\
Th/Nd & $-0.91$ & $10.29$ & $-0.928$ & $9.45$&$5.99$\\
Th/Sm & $-0.61$ & $8.89$ & $-0.796$ & $0.19$&$5.23$  \\
Th/Eu & $-0.33$ &6.08& $-0.240$ & $10.09$&$6.29$ \\
Th/Gd & $-0.81$ & $7.02$ & $-0.569$ & $18.29$&$6.29$ \\
Th/Tb & $-0.12$ & $6.08$ & \nodata & \nodata & 5.71\\
Th/Dy & $-0.89$ & $8.89$& $-0.827$ & $11.84$&$5.46$ \\
Th/Ho & \nodata & \nodata & $-0.017$ & $12.77$&$5.71$ \\
Th/Er & $-0.68$ & $6.08$ & $-0.592$ & $10.29$&$6.29$\\
Th/Tm & $0.12$ & $6.55$ & $0.155$ & $8.19$&$8.05$\\
\hline
Average & & $6.93\pm2.64$ & & $11.96\pm 4.00$ & $6.1$ \\ 
\hline \hline
\multicolumn{6}{c}{J2005$-$3057} \\
\hline
Th/Ba & \nodata & \nodata & $-1.058$ & $18.80$&$5.23$\\
Th/La & $-0.60$ & $6.55$ & $-0.362$ & $17.66$&$6.95$\\
Th/Ce & $-0.79$ & $9.36$ & $-0.724$ & $12.44$&$6.95$ \\
Th/Pr & $-0.30$ & $11.69$ & $-0.313$ & $11.09$&$5.71$\\
Th/Nd & $-0.91$ & $11.23$ & $-0.928$ & $13.56$&$6.95$\\
Th/Sm & $-0.61$ & $14.03$ & $-0.796$ & $5.33$&$6.29$\\
Th/Eu & $-0.33$ & $10.86$ & $-0.240$ & $14.97$&$5.71$\\
Th/Gd & $-0.81$ & $9.82$ & $-0.569$ & $21.10$&$5.99$\\
Th/Dy & $-0.89$ & $11.69$& $-0.827$ & $17.91$& $6.62$\\
Th/Ho &\nodata &\nodata & $-0.017$ & $18.85$ & $7.31$\\
Th/Er & $-0.68$ & $12.16$ & $-0.592$ & $16.72$ & $9.32$\\
Th/Tm & $0.12$ & $4.68$ & $0.155$ & $6.31$ & $5.23$\\
\hline
Average  & & $10.21\pm2.77$& & $15.40\pm4.27$ & 6.5 \\
\enddata \label{table:ages}
\tablenotetext{i}{ Production ratios from \citet{schatz02}.} \tablenotetext{ii}{ Production ratios from \citet{hill17}.}
\tablecomments{Age averages are given $\pm$ the standard deviation of the age measurements included in the average, for illustrative purposes. Note that the Th/Sm initial production ratio from Hill et al. 2017 was not considered in the age averages.}
\end{deluxetable}

\acknowledgements We thank Chris Sneden for providing an up-to-date
version of his neutron-capture line lists. M.C. and M.G. acknowledge
support from the MIT UROP program.  A.F. is supported by NSF-CAREER
grant AST-1255160. This work benefited from support by the National
Science Foundation under Grant No. PHY-1430152 (JINA Center for the
Evolution of the Elements). APJ is supported by NASA through Hubble
Fellowship grant HST-HF2-51393.001 awarded by the Space Telescope
Science Institute, which is operated by the Association of
Universities for Research in Astronomy, Inc., for NASA, under contract
NAS5-26555. JM thanks support from FAPESP (2012/24392-2,
2014/18100-4). This work made extensive use of NASA's Astrophysics
Data System Bibliographic Services and the Python libraries
\texttt{numpy} \citep{numpy}, \texttt{scipy} \citep{scipy},
\texttt{matplotlib} \citep{matplotlib}, and \texttt{astropy}
\citep{astropy}.


\facilities{Magellan:Clay (MIKE), Gemini:South (GMOS), KPNO:Mayall (RC spectrograph)}
\software{MOOG~\citep{Sneden73,Sobeck11}, MIKE Carnegie Python Pipeline~\citep{Kelson03}, IRAF~\citep{irafa, irafb}, NumPy~\citep{numpy}, SciPy~\citep{scipy}, Matplotlib~\citep{matplotlib}, Astropy~\citep{astropy}}

\clearpage
\startlongtable
\begin{deluxetable*}{lrrrrcrcrrrrrrrrrrrr}
\tabletypesize{\footnotesize}
\tabletypesize{\tiny}
\tablewidth{0pc}
\tablecaption{\label{table:data}
Equivalent width and synthesis measurements for J0858$-$0809,  J1432$-$4125, and J2005$-$3057}
\tablehead{ \colhead{Element} &
\colhead{$\lambda$} & \colhead{EP} & \colhead{$\log$ \textit{gf}} & \colhead{EW 1}  & \colhead{$\log\epsilon$(X) 1} & \colhead{EW 2}  & \colhead{$\log\epsilon$(X) 2} & \colhead{EW 3}  & \colhead{$\log\epsilon$(X) 3} \\
\colhead{}& \colhead{[\AA]} & \colhead{[eV]} & \colhead{}  & \colhead{[m\AA]} &  \colhead{} & \colhead{[m\AA]} &  \colhead{} & \colhead{[m\AA]} &  \colhead{}}
\startdata
C & 4311.0 & $\cdots$ & $\cdots$ & syn & 5.27 & syn & 5.69 & syn & 5.01 \\
C & 4323.0 & $\cdots$ & $\cdots$ & syn & 5.29 & syn & 5.67 & syn & 5.04 \\
O I & 6300.3 & 0.00 & $-$9.82 & 5.5 & 6.44 & \nodata & \nodata & 8.9 & 6.55 \\
Na I & 5889.9 & 0.00 & 0.11 & 164.2 & 3.43 & 136.8 & 3.62 & 204.7 & 3.76 \\
Na I & 5895.9 & 0.00 & $-$0.19 & 149.0 & 3.47 & 121.1 & 3.61 & 179.6 & 3.73 \\
Mg I & 3829.4 & 2.71 & $-$0.21 & 165.9 & 4.95 & 147.5 & 5.01 & 181.5 & 5.00 \\
Mg I & 3832.3 & 2.71 & 0.27 & 198.8 & 4.84 & 180.0 & 4.89 & 228.7 & 4.93 \\
Mg I & 3838.3 & 2.72 & 0.49 & 219.4 & 4.80 & 208.3 & 4.89 & 250.7 & 4.86 \\
\enddata
\label{table:data}
\end{deluxetable*}

${ }$

\end{document}